\documentclass[conference]{IEEEtran}
\IEEEoverridecommandlockouts
\usepackage{amsmath,amssymb,amsfonts}
\usepackage{soul}

\usepackage[linesnumbered,ruled,vlined]{algorithm2e}
\usepackage{array}
\usepackage{float,lscape}
\usepackage[utf8]{inputenc}
\usepackage{lscape} 
\usepackage{pifont}
\usepackage{algpseudocode}
\usepackage{csquotes}
\usepackage{graphicx}
\usepackage{textcomp}
\usepackage{xcolor}
\def\BibTeX{{\rm B\kern-.05em{\sc i\kern-.025em b}\kern-.08em
    T\kern-.1667em\lower.7ex\hbox{E}\kern-.125emX}}
\begin{document}
\title{PRAGyan - Connecting the Dots in Tweets}



\author{\IEEEauthorblockN{Rahul Ravi}
\IEEEauthorblockA{\textit{Dept. of Computer Science} \\
\textit{University of Calgary}\\
rahul.ravi@ucalgary.ca}
\and
\IEEEauthorblockN{Gouri Ginde}
\IEEEauthorblockA{\textit{Dept. of Electrical and Software Engineering} \\
\textit{University of Calgary}\\
gouri.deshpande@ucalgary.ca}
\and
\IEEEauthorblockN{Jon Rokne}
\IEEEauthorblockA{\textit{Dept. of Computer Science} \\
\textit{University of Calgary}\\
rokne@ucalgary.ca}
}

\maketitle

\begin{abstract}
As social media platforms grow, understanding the underlying reasons behind events and statements becomes crucial for businesses, policymakers, and researchers.
This research explores the integration of Knowledge Graphs (KGs) with Large Language Models (LLMs) to perform causal analysis of tweets dataset. The LLM aided analysis techniques often lack depth in uncovering the causes driving observed effects. By leveraging KGs and LLMs, which encode rich semantic relationships and temporal information, this study aims to uncover the complex interplay of factors influencing causal dynamics and compare the results obtained using GPT-3.5 Turbo. We employ a Retrieval-Augmented Generation (RAG) model, utilizing a KG stored in a Neo4j (a.k.a PRAGyan) data format, to retrieve relevant context for causal reasoning. Our approach demonstrates that the KG-enhanced LLM RAG can provide improved results when compared to the baseline LLM (GPT-3.5 Turbo) model as the source corpus increases in size. Our qualitative analysis highlights the advantages of combining KGs with LLMs for improved interpretability and actionable insights, facilitating informed decision-making across various domains. Whereas, quantitative analysis using metrics such as BLEU and cosine similarity show that our approach outperforms the baseline by 10\%.
\end{abstract}


\section{Introduction}
Understanding the underlying reasons behind events and statements is of great interest for businesses, policymakers, and researchers alike. These reasons may frequently be found from available textual data. An important source is in-line social media which has become widely used and as such it has developed into a rich source for such data  \cite{deshpande2018user}. Large Language Models (LLMs) such as GPT-3.5 Turbo and BERT provide deep, context-aware language understanding, capturing contextual and temporal information critical for deciphering the nuances of social media discourse \cite{devlin2019bert}. However, these techniques often offer surface-level insights and lack the depth to uncover root causes \cite{jin2023cladder}. In this paper, Knowledge Graphs (KG), based on graph theory fundamentals, storing data as nodes and links, offering structured representations of domain-specific knowledge and capturing rich semantic relationships are considered for data representation \cite{kejriwal2021knowledge}, \cite{dieudonat2020exploring}, \cite{peng2023knowledge}. Using this representation temporal information, which is also crucial for understanding causal relationships in social media data, is used since the sequence and timing of events provide essential insights into causation \cite{hong2017event2vec}. \textbf{Our aim is therefore} to extract causal information from social media content, leveraging both KGs and LLMs with temporal knowledge for enhanced causal analysis.

Thus, in this research, we explore the effectiveness of integrating knowledge graphs (KGs) and large language models (LLMs) in performing causal analysis of text-based social media data such as tweets from the X (formerly Twitter) platform \cite{pan2024unifying}. We hypothesize that these two approaches (LLM and KG) can enable an effective detection of causes and their underlying effects, providing a richer understanding of the dynamics in social media interactions \cite{dieudonat2020exploring}. Thus, \textbf{in this study} we explore and compare the hybrid method, a combination of KG and LLM, to enhance the interpretability and actionable insights derived from causal analysis by delving deeper into the underlying drivers of events and statements in time series tweets. As such, we compare the GPT model with our hybrid approach PRAGyan for causal analysis on the COVID-19 tweets dataset.
\textbf{Hence, we propose PRAGyan\footnote{PRAGyan means wisdom and gyan means knowledge in Sanskrit language}}, a hybrid KG-RAG-LLM model with Retrieval-Augmented Generation (RAG) as a key component, binding KG and LLM together. RAG combines retrieval mechanisms with generative models to enhance the generation process by providing relevant contextual information. Taking inspiration from \cite{ding2024survey}, we use RAG as a retriever model to fetch pertinent documents or knowledge snippets along with a generator model to produce responses conditioned on both the query and the retrieved information. This ensures contextually accurate and informative outputs, leveraging the vast knowledge available from the KGs together with the sophisticated language understanding acquired from using LLMs \cite{gao2023retrieval}. This dynamic integration makes RAG particularly well-suited for tasks requiring up-to-date information extraction and complex reasoning, such as causal analysis acquired from social media data \cite{lewis2020retrieval}.

\noindent \textbf{Our main contributions in this study are:}

\begin{itemize}
    \item We propose a novel methodology to derive actionable insights from a time series analysis of a COVID-19 tweets dataset \cite{loureiro2022timelms} to derive deeper insights to events for causal analysis. 
    \item Our proposed approach can trace back the list of tweets which lead to such causations thus bringing in interoperability and transparency which is currently missing from LLM only solutions (such as GPT models) \cite{kiciman2023causal, jin2023can}.
    \item Our method can be used on a continuously growing dataset and not just on a snapshot of a dataset.
    \item We evaluate our method using both qualitative and quantitative techinques.
\end{itemize}

The paper is structured as follows: Section \ref{motivation} sets the context with a motivating example. Section \ref{Background} details various elements of this study. Section \ref{Methodology} explains the details of the usage of the methodology and dataset. Section \ref{Results} showcases the main results and  Section \ref{RelatedWork} captures related work. Section \ref{TTV} lists the limitations of the study and Section \ref{conclusion} highlights future works and conclusion.  



\section{Motivating example}
\label{motivation}
During the COVID-19 pandemic, we noticed a recurring theme in tweets where people frequently mentioned an increased appreciation for simple things in life. This observation sparked our curiosity, and we attempted to manually understand the reasons behind this sentiment. However, this process proved to be extremely challenging due to the scattered and unstructured nature of the relevant tweets.

To answer a question like, \enquote{What caused people to appreciate simple things around them during the COVID-19 pandemic?} we had to sift through countless tweets. These tweets were not posted sequentially nor were they directly related to each other. The information was dispersed across different parts of the dataset, making it difficult to piece together a comprehensive answer manually.

After spending significant manual effort we found that people’s appreciation for simple things was influenced by various interconnected factors. The lockdown measures led to isolation, which in turn heightened people's awareness of their immediate surroundings. The restriction of activities forced individuals to spend more time at home, allowing them to notice and value the simple things that were always around them. This combination of factors provided a sense of comfort and appreciation during challenging times.

This manual process highlighted the need for a more efficient solution. The difficulty in manually analyzing the vast and scattered data from social media underscored the necessity of developing a method that could automatically integrate and analyze this information to uncover deeper insights. Thus, we aimed to understand and interpret the complex dynamics of social media interactions represented on a temporal graph and automate this difficult task further. 

\section{Key Concepts}
\label{Background}
In this section, we discuss the essential preliminaries of our research. \\
\textbf{Knowledge Graphs (KGs):}
KGs provide semantic frameworks for organizing domain-specific information, facilitating causal analysis of social media data \cite{peng2023knowledge, dieudonat2020exploring, kejriwal2021knowledge, xu2023data}. By enriching textual data with structured semantic information, KGs enhance context and relationship understanding, aiding in identifying underlying drivers of events and causal relationships. They encode entity relationships and attributes, offering deeper insights, in particular, into social media discussions and integrating diverse information sources, thereby improving the robustness of causal analysis \cite{dessi2021generating}.
Neo4j, which is a leading graph database platform, excels in managing complex relationships within large text corpora, including temporal information on edges, which allows accurate representation of event chronology and causal links \cite{tjortjisgraph}. Its flexible schema and query language make it ideal for dynamic, heterogeneous data like social media content, enabling timely analysis and sophisticated queries for deeper insights. \\
\textbf{Large Language Models:}
LLMs such as GPT and BERT have revolutionized natural language understanding tasks \cite{minaee2024large}. Trained on vast corpora of textual data, LLMs capture intricate semantic nuances within text, enabling reasoning and inference tasks \cite{devlin2019bert}. GPT-3.5 Turbo, an advanced version of GPT models by OpenAI, is used as the baseline in this study. It excels in comprehending queries and context, producing accurate and contextually relevant responses \cite{zeidisynergizing}. Its ability to handle longer and complex contexts is crucial for our study, and it is accessible through OpenAI's API for high-quality causal reasoning. \\
\textbf{Embedding and Encoding Techniques:}
Dimension reduction transforms high-level representations of text or graph into a lower dimensional 2D  space as vectors \cite{van2009dimensionality}. When performed on a graph with a list of nodes and edges, it is referred to as \enquote{embedding} and when performed on a sequential text, it's known as \enquote{encoding}. This research employs both KG and LLM for optimized context extraction: \\
- \textit{Node2Vec:}
Node2Vec generates node embeddings by simulating random walks on the graph and applying the Skip-Gram model, capturing semantic similarities based on structural context \cite{wang2021kg2vec} \cite{grover2016node2vec}. These embeddings capture semantic and temporal relationships between nodes, essential for understanding connections in social media data. They facilitate the identification of underlying factors driving events and statements, enabling efficient transfer learning and informed decision-making. \\
- \textit{Sentence-BERT: }
BERT (Bidirectional Encoder Representations from Transformers) is used for text comprehension \cite{wang2024utilizing} \cite{reimers2019sentencebert}. Input data is converted lowercase and tokenized it into subword units before generating encodings by summing token, segment, and position encodings. Final encodings for each token are then obtained based on the NLP task. 
\\ 
\textbf{LLM-Enhanced KG:}
The process begins with a fine-tuned relation extraction LLM (LLAMA3 \cite{towardsDS}), which identifies entities and relationships from text to generate triples which are used to construct a KG. This integration enhances the ability to identify and articulate the intricate web of correlations between events and statements, allowing for a more structured exploration of complex interactions within the data. \\
\textbf{Retrieval-Augmented Generation (RAG):}
The KG, in conjunction with Sentence-BERT LLM is used to retrieve relevant context through Retrieval-Augmented Generation (RAG). After the graph has been embedded, the LLM is used to identify contextually similar subsets of corresponding nodes and edges with query encoding, which are then expanded based on the graph structure for obtaining both contextually and semantically relevant context. \\
\textbf{KG-Enhanced LLM:}
Finally, the retrieved context is used by GPT-3.5 Turbo to reason about the causes behind user-defined queries. Traditional language models like GPT excel in generating context-aware responses but often lack the structured knowledge to fully grasp complex relationships in social media data \cite{chan2024exploring}. Hence, the context enhances the model's responses with structured correlation knowledge from the KG. This integration improves the model's ability to identify and analyze causal relationships, improving the accuracy and depth of causal analysis.

The combined approach outperforms using standalone LLMs (the baseline) or KGs, leveraging the strengths of both techniques. The upcoming sections discusses an implementation strategy for this approach, in PRAGyan, and the results highlight the efficiency and effectiveness of this integrated method in uncovering causal factors driving events and statements on social media platforms through comparative analysis.


\section{Methodology}
\label{Methodology}
\subsection{Research Question}
We have formulated a key research question focused on evaluating the effectiveness of combining large language models (LLMs) with knowledge graph (KG) embeddings for causal analysis:

\textbf{Does the integration of KG and LLM through RAG enhance causal reasoning capabilities on social media data as opposed to using LLMs without context?}

\textit{Explanation:} This question investigates whether integrating knowledge graph embeddings with LLMs enhances the context and accuracy of causal reasoning in social media data. By combining the structured knowledge in KGs with the deep contextual understanding of LLMs, we hypothesize that this approach will yield superior results for identifying and reasoning about causation. The expected inputs include raw social media text data, knowledge graph embeddings, and LLMs for encoding and prompt engineering. The goal is to demonstrate improved performance in causal reasoning tasks using the combined approach compared to using LLMs alone. The study design that follows encompasses key steps involved in creation of the proposed hybrid model (PRAGyan) and to answer our RQ.

\subsection{Dataset}
X (formerly Twitter) provides rich source of first hand experience details and information reporting as it is frequently used by lay people and hence it is one of the primary platforms for individuals where they share real-time updates, personal experiences, and opinions on various topics. We obtained the COVID19 Tweets dataset from Kaggle \cite{gabriel_preda_2020}. The dataset had thirteen features/fields. However, we only used 
the \enquote{text} that contains all the tweets along with \enquote{date} that contains timestamps for each tweet.Future work might include \enquote{user\_location} (geography based analysis of causes) and \enquote{date} (temporal parameter that directly affects causation). 
This is a popular dataset with more than 22k downloads to date, mainly for data cleaning and learning purposes as stated on the website \cite{gabriel_preda_2020}.

The tweets were gathered between 02/28/2020 and 07/23/2020 with a coverage across the world using Twitter API with the hashtag, \enquote{\#covid19}. It's size is 68.71MB and contains over 35k rows. On average each tweet has around 17 words after pre-processing.

\subsection{Study design} 
Our study design comprises two main sections: the baseline approach and the proposed model. Each section tackles the problem of causal reasoning in social media data, specifically focusing on tweets, but they employ different methodologies. \\
\begin{figure*}[!htpb]
    \centering
\includegraphics[scale=.18]{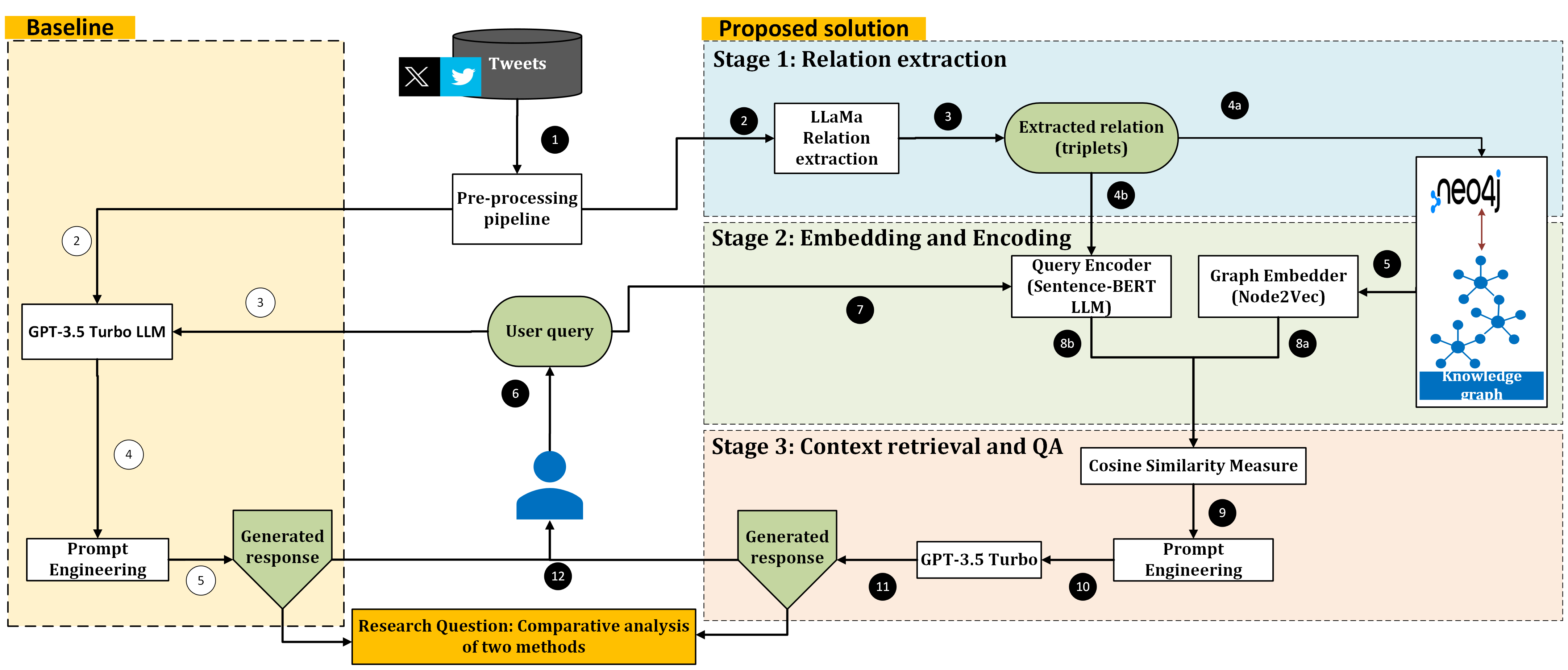}
    \caption{ Our study design: On the left, the baseline approach and corresponding workflow is depicted.  On the right, our proposed model: PRAGyan and its corresponding workflow is depicted. Various elements in the middle section are common to both the methods - Steps \ding{182}, \ding{187} \& 12}
    \label{fig:StudyDesing}
\end{figure*}
\textbf{Pre-processing:}
Figure \ref{fig:StudyDesing} shows the overall architecture of our study.  The first step (\ding{182}) is the data pre-processing which is common to both the baseline method and our proposed method. In addition to the tweets' text, we also extract timestamps from tweets and ensure their accuracy by handling different time zones, correcting any inconsistencies, and standardizing the format.
The tweets strings are highly unstructured, ambiguous and full of noise (special characters, acronyms and URL's), Thus we first utilize a dictionary to expand contractions of words often used in social media slang replacing them with their appropriate representation. For example, \enquote{aren't}: \enquote{are not / am not}, \enquote{can't've}: \enquote{cannot have}, etc. 
We also remove special characters, emojis, URLs and Hashtags. 

\subsubsection{The Baseline:} The baseline model is using the GPT-3.5 Turbo LLM directly for querying the entire preprocessed text corpus (Step \ding{173}) through prompt engineering for reasoning causation (Step \ding{175}) when a user query is passed to it (Step \ding{174}) and result is generated in Step \ding{176}. Unlike the proposed model, this method does not utilize a knowledge graph for context retrieval. Instead, it relies solely on the LLM's ability to process and generate responses based on the raw text data.

\subsubsection{Proposed Model (PRAGyan)}
The proposed PRAGyan model follows a three-stage process: Relation extraction, Embedding/Encoding, and Context Retrieval and QA, as illustrated in Figure \ref{fig:StudyDesing}. Each stage leverages specific technologies chosen for their capabilities to enhance the overall process.\\ \\
\textbf{Relation Extraction:}
In the proposed model, relation extraction is performed using the LLaMa3 model (Step \ding{183} in the diagram), specifically the fine-tuned variant from Hugging Face (solanaO/llama3-8b-sft-qlora-re \cite{towardsDS}). This model identifies entities and their relations (together known "triplets") from the pre-processed tweets (Step \ding{184}). LLaMa3 \cite{wadhwa2023revisiting} enhances traditional relation extraction methods through supervised fine-tuning by leveraging advanced natural language understanding. The process begins by loading the pre-trained model with a PEFT adapter for efficient memory usage. The tokenizer creates prompts from the preprocessed text, and the model generates outputs, forming the triples. They are stored as a KG within a graph database, Neo4j (Step \ding{185}a). Subsequently, the Sentence-BERT encodings are also obtained for all the relational triples (Step \ding{185}b) which are stored as edge property information in the database. \\
\textbf{Embedding and Encoding:}
In this stage, first the KG stored in Neo4j is embedded using Node2Vec (Step \ding{186}). This method captures both local and global structural information of the graph, creating high-quality vector representations stored in Neo4j for efficient retrieval of contextually relevant data. 
The objective is to maximize the probability of observing a node’s neighbors given its embedding through the Node2Vec algorithm:
\[
\max \sum_{u \in V} \sum_{v \in N(u)} \log \Pr(v|u)
\]
where \(\Pr(v|u)\) is the probability of node \(v\) being a neighbor of node \(u\), computed using the softmax function:
\[
\Pr(v|u) = \frac{\exp(\mathbf{z}_v^\top \mathbf{z}_u)}{\sum_{w \in V} \exp(\mathbf{z}_w^\top \mathbf{z}_u)}
\]
Here, \(\mathbf{z}_u\) and \(\mathbf{z}_v\) are the embeddings of nodes \(u\) and \(v\), respectively. The learned embeddings are stored in the graph database for efficient retrieval and analysis. In a dynamic graph situation where new events continuously occur, it is running periodically while incorporating the latest temporal data ensures the embeddings stay up-to-date with the evolving graph.

When a user inputs a query (Step \ding{187}) to reason causation, the query is encoded using an LLM (Sentence-BERT) to create a vector representation (Step \ding{188}). This representation is mapped onto the corresponding Node2Vec embeddings to identify the most relevant context in the next stage. \\ \\
\textbf{Context Retrieval and QA:}
This step involves retrieving the most relevant contexts from the knowledge graph to engineer prompts, a process known as prompt engineering (Step \ding{190}). These prompts are fed into a GPT-3.5 Turbo model (Step \ding{191}) to generate a response. The response, enriched with relevant context from the knowledge graph, is provided to the user, ensuring a detailed and comprehensive answer. This is known as RAG. It involves:
\begin{itemize}
    \item \textit{Identifying Contextually Relative Subgraph}: The Sentence-BERT encoding of the query is compared with the encodings calculated from the triples to choose a highly similar set through Cosine similarity (Step \ding{189}a). This set is then looked up on the Node2Vec embedding space with their corresponding embeddings. The resulting set of node embeddings are contextually similar to the query (Step \ding{189}b) which are passed on to the next step for further refinement to extract contextual information.
    \item \textit{Retrieving Final Context - Contextually and Semantically Relevant}: With this set, using Cosine similarity, we identify it's top semantically similar subsets of neighbouring {nodes, edges} in the Node2Vec embedding space (Step \ding{190}). Both similarity measures (this and previous step) are validated for a threshold of 0.35, which was determined through trials and inspections of contexts with various thresholds. This ensures the right amount of context that could potentially provide the true cause. The subgraphs are mapped onto the preprocessed sentences, retrieving the top-ranked sentences (contextually and semantically most similar to the input statement) that serve as the source context for prompting with LLM (Step \ding{191}). Including temporal constraints ensures the retrieved context respects the chronological order of events.
\end{itemize}
The retrieved context is then used by the GPT-3.5 Turbo LLM to generate responses with an effective prompt instructing the model to infer the cause for the input query (Step \textbf{11}). In the last step (Step \textbf{12}) a comparative analysis of the two methods is then performed.




\subsection{Evaluation and Comparative Analysis}
The performance of the integrated approach (PRAGyan) is evaluated and compared with the baseline approach (GPT3.5 Turbo). 
The evaluation process is conducted as follows:
\subsubsection{Qualitative evaluation}
According to \cite{gao2023retrieval}, qualitative metrics like faithfulness, answer relevance, and context relevance provide a nuanced understanding of how well the models capture and convey meaningful information. The paper \cite{es2023ragas} further emphasizes the importance of automated qualitative evaluation, highlighting challenges such as ensuring consistency and aligning automated metrics with human judgment. These qualitative assessments are crucial for determining the effectiveness of models for generating accurate and contextually relevant responses, which is vital for the success of our research.

The baseline processes the entire corpus without specific context from the knowledge graph, generating responses based solely on the raw text data. The evaluation process involves manually designed queries from the tweets and correct causes created for validation. The goal is to determine which approach yields the closest truth or most relevant cause statements to the right answers based on specific metrics.
The baseline is not fine-tuned for this specific task; instead, it utilizes the pre-trained model to generate responses based on the provided context.
\subsubsection{Quantitative evaluation}
Quantitative evaluation is performed utilizing two metrics as follows:
\begin{itemize}
    \item \textbf{BiLingual Evaluation Understudy (BLEU) Score}:  BLEU measures the precision of n-grams in generated text against reference texts, evaluating word or phrase matches with reference causes \cite{papineni2002bleu}. It's simple, easy to understand, and reliable for large-scale evaluations. Although it doesn't capture word importance or synonyms, BLEU provides a quick, reproducible measure of text accuracy, which is crucial for assessing if the model generates causes close to the ground truth \cite{reiter2018structured}.
    \item \textbf{Cosine Similarity on Sentence-BERT Encodings}: The BERT Similarity metric, based on pre-trained BERT embeddings, captures contextual meanings and semantic similarity between texts, making it ideal for assessing the conveyed information rather than exact structure \cite{reimers2019sentencebert}. SentenceBERT, optimized for sentence embeddings, shows a high correlation with human judgments of similarity, surpassing BLEU and better reflecting the quality and relevance of generated summaries \cite{choi2021evaluation}. The similarity score ranges from 0 to 1, with 1 being the highest.
    
\end{itemize}

Together, these metrics ensure a comprehensive evaluation of both structural and semantic quality in generated text, as highlighted in \cite{haque2022semantic}. Hence, they are well suited for evaluating performance for this use case.

\section{Results}
\label{Results}

\begin{table}[!htbp]
\renewcommand{\arraystretch}{1.2}
\caption{Evaluation Metrics for the Baseline and PRAGyan Models}
\begin{center}
\begin{tabular}{c|c c}
 
\textbf{Metric} & \textbf{The Baseline} & \textbf{PRAGyan} \\
\hline
\textbf{BLEU} & 0.45 & \textbf{0.49}\\
 
\textbf{Cosine Sim with LLM Encoder} & 0.86 & \textbf{0.88} \\

\end{tabular}
\label{tab1}
\end{center}
\end{table}
To create the proposed model, we used Google Colab Pro equipped with A100 GPU and a high RAM setting of about 80GB and Python 3.11 as the programming environment. We also utilized OpenAI APIs, and libraries from Huggingface, NLTK, SkLearn packages \cite{wolf2020transformers} for generating the results.
\subsection{Quantitative evaluation}
\begin{figure}[!htpb]
    \centering
    \includegraphics[trim={0 0 0 .68cm},clip,scale=.25]{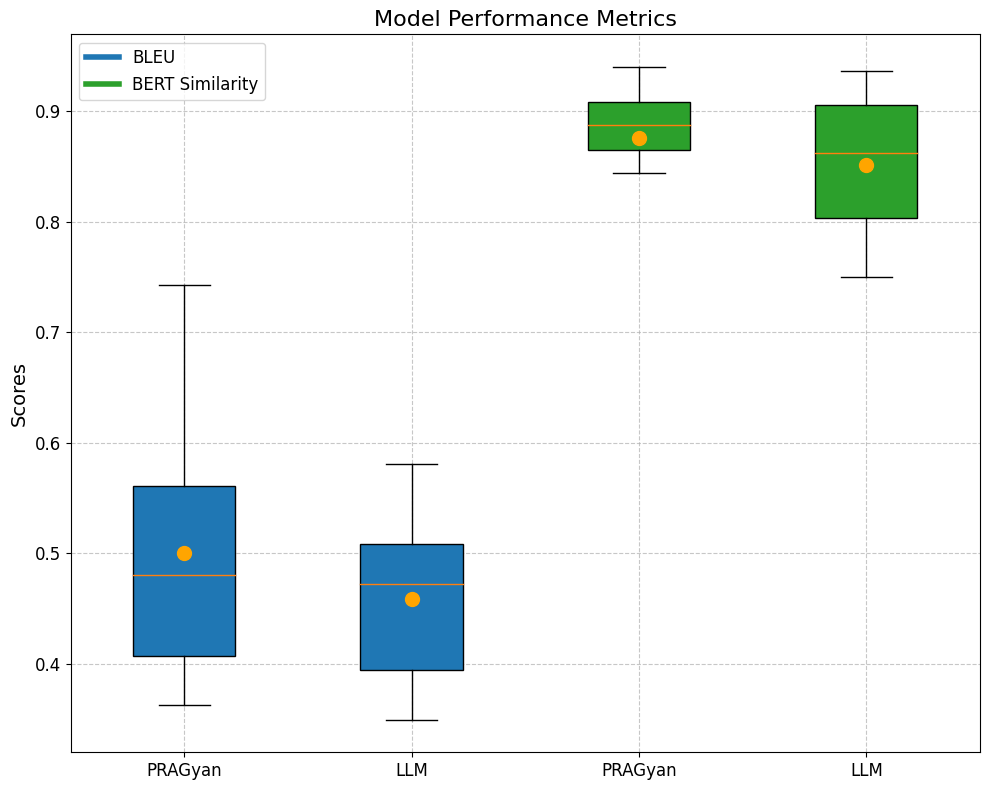}
    \caption{Box plot showing metrics between the Baseline and PRAGyan}
    \label{fig:boxPlot}
\end{figure}

The two average metric scores for both models observed over multiple queries are presented in Table \ref{tab1}. PRAGyan outperforms the baseline model across both the metrics, indicating its superior performance in generating responses that are both accurate and comprehensive.
Based on the box plots in Figure \ref{fig:boxPlot}, we can infer the following about the performance of the two models:
\begin{enumerate}
    \item  \textbf{BLEU score} \\
- \textit{PRAGyan}: It shows a wider range of BLEU scores, with some instances achieving very high scores, suggesting the model's capability to produce highly relevant and accurate text in some cases. The higher variability, indicated by the wide interquartile range and the position of the average score above the median, suggests that while the model can perform very well, it is not consistently high-performing across all instances. \\
- \textit{The Baseline}: The BLEU scores for the baseline are more tightly clustered, indicating more consistent performance. However, the lower average and median scores suggest that, on average, this model does not achieve the same high level of relevance and accuracy as PRAGyan. The consistency is beneficial for tasks requiring stable performance but may lack the peak performance seen in PRAGyan. 

\item \textbf{Cosine similarity} \\ 
- \textit{PRAGyan}: The BERT Encoding Cosine Similarity scores for PRAGyan are consistently high, with a median score of approximately 0.89. This indicates that at least half of the similarity scores are at or above this value, highlighting the model’s effectiveness in capturing semantic relationships. The small interquartile range (IQR) suggests that the scores are tightly clustered around the median, demonstrating reliable performance in terms of semantic similarity with minimal variation.\\
- \textit{The Baseline}: The BERT Encoding Cosine Similarity scores for the baseline model are generally lower compared to PRAGyan, with a median score of approximately 0.86. The wider range and greater variation in scores suggest that the baseline model struggles more with certain instances, leading to less consistent and reliable performance in capturing semantic relationships. \\ 
\end{enumerate}
Our proposed model, leverages the strengths of both KG and RAG, resulting in better overall performance. It showed significant advantages for the COVID-19 tweets dataset, generating contextually and semantically relevant and accurate information, thus, improving the quality of insights derived from this social media content. However, higher score variations suggest room for improvement in the knowledge graph’s construction, as it currently only conveys correlation information.
In contrast, while the baseline provides more consistent performance, its generally lower scores indicate less accuracy. As the data environment grows, this increases the overhead for the baseline, potentially affecting results and processing times. Hence, maintaining a graph database helps with dynamic real-time social media updates and faster querying.\\

\noindent\fcolorbox{black}{lightgray}{
    \parbox{.98\linewidth}{
PRAGyan demonstrates superior average performance and the capability to produce highly relevant and accurate text, as indicated by higher metric scores. This makes it particularly beneficial for analyzing social media data, where context and relevance are crucial. The variability in its performance highlights its potential to excel in many instances, suggesting that integrating LLM with KG through RAG enhances the model's ability to generate high-quality text.}} \\
\subsection{Qualitative evaluation
}
Table \ref{tab2} illustrates a comparative analysis of the responses generated by PRAgyan and the baseline in addressing questions about the COVID-19 pandemic. 


For the first question regarding measures to support COVID-19 patients, PRAGyan identifies specific actions such as the setup of a plasma bank, prayer support, and updates on ICU patients, highlighting concrete steps taken to enhance treatment outcomes. In contrast, the baseline provides a general summary, lacking specific details.

Similarly, for the second question about the impact of the pandemic on grocery shopping behaviors and store operations, PRAGyan details instances of panic buying, dedicated shopping hours for the elderly, and increased demand for online shopping. It also mentions specific challenges faced by stores, like shortages of essential items and the pressure on employees to keep shelves stocked. On the other hand, the baseline gives a more generalized response, focusing broadly on fear of shortages and stockpiling without specific contextual examples.
\\  \\ \noindent\fcolorbox{black}{lightgray}{
    \parbox{.98\linewidth}{
 The baseline model generates responses based solely on the raw text data, leading to variations in responses for the same set of queries when run multiple times. This is a characteristic of generative AI models like LLMs, which can produce different outputs depending on the input context and randomness inherent in the generation process, unless fine tuned comprehensively for datasets or most efficient prompt engineering. In contrast, PRAGyan enhances the generation process by providing specific context derived from the knowledge graph, resulting in more consistent and accurate responses.}} \\

This comparison underscores the enhanced ability of PRAGyan to provide richer, context-aware responses by utilizing structured knowledge from the KG, thereby offering deeper insights and more actionable information.
    \begin{table*}[!htpb]
    \centering
    \caption{Responses from PRAGyan (Context from Embeddings through Selective Tweets) and the Baseline (Entire Corpus as the Text Source)}
\begin{tabular}{|p{2cm}|p{3cm}|p{3cm}|p{6cm}|}
\hline
\textbf{Question} & \textbf{Summary from Baseline} & \textbf{Summary with Context (PRAGyan)} & \textbf{Tweets} \\ \hline
What measures are being taken to support patients recovering from COVID-19? & Measures are being taken to support patients recovering from COVID-19 due to the impacts of the pandemic and the need to address the health and well-being of individuals affected by the virus. & Efforts to enhance treatment, such as the setup of a plasma bank, prayer support, and sharing of updates on ICU patients indicate a focus on improving recovery outcomes for COVID-19 patients & \begin{tabular}[p{6cm}]{@{}p{6cm}@{}}{[}2020-07-25{]} {[}12:21 PM{]} Rajasthan government today started plasma bank at sawai man singh hospital in jaipur for treatment of covid 19. \\ {[}2020-07-25{]} {[}12:25 PM{]} Mom is in icu due to covid just want prayer from you and everyone who is listening you covid19. \\ {[}2020-07-25{]} {[}12:27 PM{]} Praying for good health and recovery from covid19.\end{tabular} \\ \hline
How did the COVID-19 pandemic affect grocery shopping behaviors and store operations? & Fear of shortages and uncertainty due to the COVID-19 pandemic led to panic buying and stockpiling of essential items, such as food, toilet paper, and sanitizing products, in anticipation of potential disruptions in supply chains and daily routines. & The COVID-19 pandemic led to panic buying, resulting in empty shelves and increased demand for online shopping. Grocery stores implemented measures like dedicating shopping hours for the elderly and encouraging online orders. The pandemic caused shortages of essential items like toilet paper and hand sanitizer, and store employees faced increased pressure to keep shelves stocked. & \begin{tabular}[m{6cm}]{@{}m{6cm}@{}}{[}2020-03-15{]} {[}09:00 AM{]} Coronavirus Australia: Woolworths to give elderly, disabled dedicated shopping hours amid COVID-19 outbreak  \\ {[}2020-03-16{]} {[}09:25 AM{]} Due to the Covid-19 situation, we have increased demand for all food products... The wait time may be longer for all online orders, particularly beef share and freezer packs. \\ {[}2020-03-16{]} {[}10:00 AM{]} As news of the region’s first confirmed COVID-19 case came out of Sullivan County last week, people flocked to area stores to purchase cleaning supplies, hand sanitizer, food, toilet paper and other goods. \\ {[}2020-03-16{]} {[}11:00 AM{]} All month there hasn’t been crowding in the supermarkets or restaurants, however, reducing all the hours and closing the malls means everyone is now using the same entrance and dependent on a single supermarket\end{tabular} \\ \hline
\end{tabular}
\end{table*}

\section{Related Work}
\label{RelatedWork}
The growing interest in computational methods for analyzing dynamic text data, such as social media, has led to significant advancements in data representation and analysis techniques, with KGs and LLMs at the forefront.\\
\textbf{Advanced Graph Techniques and NER in Social Media:}
\cite{dieudonat2020exploring} explore combining contextual word encodings with KG embeddings, enhancing sentiment analysis through integrated representations. Effective Named Entity Recognition (NER) in social media texts, detailed by \cite{liu2011recognizing} and \cite{liu2013named}, informs our preprocessing steps. Additionally, a fine-tuned LLAMA3 model for relation extraction (triples) demonstrates efficiency in training on data \cite{towardsDS}.\\
\textbf{Causal Knowledge Graphs and Explainability:}
\cite{tan2023constructing}, \cite{jaimini2022causalkg}, and \cite{gopalakrishnan2023text} delve into constructing causal KGs from varied data sources and enhancing explainability through causal reasoning. These studies provide a framework for our work in creating KGs that not only predict relations and similarities but also causation, offering actionable insights into the factors driving information on social media.\\
\textbf{Causality with LLMs:}
\cite{kiciman2023causal} and \cite{jin2023can} explore the capabilities of LLMs in causal reasoning.  \cite{loureiro2022timelms} introduce TimeLMs, which are diachronic language models that adapt to the evolving linguistic landscape of social media, enhancing our ability to capture and analyze the nuances of cause-effect relationship over time.\\
\textbf{RAG Process:}
The RAG process plays a pivotal role in our methodology by using LLMs and KGs to rank similarities between input text and a list of sentences. This approach provides the top 5 ranked sentences containing causal information, thus reducing the context for further LLM-based cause extraction. \cite{pan2024unifying} propose frameworks for unifying LLMs and KGs, which we build upon to enhance predictive power and semantic richness.\\
\textbf{Combining LLMs and KGs}
The integration of KGs with LLMs signifies a transformative shift in computational methods. \cite{pan2024unifying} outline a comprehensive roadmap for merging LLMs and KGs, which we build upon. This unification harnesses the predictive power of LLMs and the rich semantic structures of KGs, overcoming limitations found in each model when used in isolation. Our research leverages these frameworks to construct an ideal pipeline that better suits our goal.

Through the integration of these pioneering studies, our research leverages cutting-edge technologies and methodologies to advance the field of causal reasoning.

\section{Limitations}
\label{TTV}
Models like Node2Vec, Sentence-BERT, and GPT-3.5 Turbo can present challenges when working with other domains, differently formatted data and working environment. They were chosen for their accessibility, and compatibility with the existing resources and environment, making them practical for this research.

Evaluating RAG and LLM models is complex due to their hybrid nature. RAG systems need assessment of both retrieval and generation components, focusing on dynamic knowledge bases and subjective measures of relevance and coherence as explained in \cite{yu2024evaluation}. Similarly, \cite{guo2023evaluating} highlights that evaluating LLMs demands a comprehensive approach to address data leakage, inappropriate content, and safety standards. 

Metrics like BLEU and Cosine Similarity provide quantitative assessments but may not fully capture qualitative aspects of causal reasoning, necessitating human evaluation. Tailoring the framework with appropriate models and metrics ensures robust and contextually relevant evaluation across different datasets and use cases.


\section{Conclusion and Future work}
\label{conclusion}
In this research we demonstrated the effectiveness of integrating Knowledge Graphs (KGs) with Large Language Models (LLMs) for causal analysis of social media data. The combined approach significantly enhances the depth and accuracy of understanding the factors driving social media interactions.

Our proposed model has shown interesting results for the COVID19 dataset, and we believe that this approach can be further used to pinpoint a wide variety of information in the tweets dataset specific to a problem. For example:
1) Such an approach might be useful to identify misinformation and the tweets that could be giving rise to it in a region or country.
2) It could be useful for identifying the most sought after development woes by the citizens based on the tweets mined for particular hashtags
3) Our approach relies on a database which forms the crux of the solution, thus, it allows us to identify original tweets which could be a problematic one in election or sensitive times such as riots and wars.

The RAG method used in our framework improves the quality of causal inferences, as validated by higher similarity scores compared to the baseline model. This integration leverages the strengths of both KGs and LLMs, providing detailed and contextually enriched insights that are crucial for informed decision-making. The added layer of temporal information further ensures that only temporally plausible relationships are considered, and it enriches the context for queries by capturing the dynamic evolution of events over time. Overall, the integration of KGs with LLMs offers a powerful methodology for analyzing the complex dynamics of social media data, yielding more precise and actionable insights.

In the future, we aim to refine the construction of the KG further by including causation information along with correlation to ensure better retrieval of context for the LLM query. This enhancement will provide a more robust understanding of causal relationships in the data. 

Additionally, we will explore alternative embedding techniques such as GraphSAGE or GAT (Graph Attention Networks) to capture richer semantic information \cite{hamilton2017inductive, velickovic2017graph}. 


\bibliographystyle{plain}
\bibliography{reference}

\end{document}